# Crossing the Tepper Line: An Emerging Ontology for Describing the Dynamic Sociality of Embodied AI

**KATIE SEABORN**
*Tokyo Institute of Technology*
*RIKEN Center for Advanced Intelligence Project (AIP)*

**PETER PENNEFATHER**
*gDial, Inc.*

**NORIHISA P. MIYAKE**
*RIKEN Center for Advanced Intelligence Project (AIP)*

**MIHOKO OTAKE-MASTUURA**
*RIKEN Center for Advanced Intelligence Project (AIP)*

# Crossing the Tepper Line: An Emerging Ontology for Describing the Dynamic Sociality of Embodied AI

**Katie Seaborn***
Tokyo Institute of Technology
Tokyo, Japan
seaborn.k.aa@m.titech.ac.jp

**Peter Pennefather**
gDial, Inc.
Toronto, Ontario, Canada
p.pennefather@gmail.com

**Norihisa P. Miyake, Mihoko Otake-Matsuura**
RIKEN Center for Advanced Intelligence Project.
Tokyo, Japan
{ norihisa.miyake, mihoko.otake }@riken.jp

## ABSTRACT

Artificial intelligences (AI) are increasingly being embodied and embedded in the world to carry out tasks and support decision-making with and for people. Robots, recommender systems, voice assistants, virtual humans—do these disparate types of embodied AI have something in common? Here we show how they can manifest as "socially embodied AI." We define this as the state that embodied AI "circumstantially" take on within interactive contexts when perceived as both social and agentic by people. We offer a working ontology that describes how embodied AI can dynamically transition into socially embodied AI. We propose an ontological heuristic for describing the threshold: the Tepper line. We reinforce our theoretical work with expert insights from a card sort workshop. We end with two case studies to illustrate the dynamic and contextual nature of this heuristic.

## 1 INTRODUCTION

Indulge us for a moment: When you imagine an artificial intelligence (AI) interacting with people, what do you see? How about when AI interact with each other? The results of this thought experiments will vary across time, culture, and circumstance. For many, AI will have a physical or virtual form (i.e., a perceivable morphology) and be interactive, sensing the environment (including other agents within it, like people) and (re)acting through various actuators [36]. For scholars and practitioners, visions of how AI are embodied during interactions with people are likely to coincide with our discipline, background, and training. Roboticists may imagine a robot, such as the popular Nao robot. Web developers may visualize a chatbot or recommender system. Game designers may think of a non-playable character or personified help system. Smartphone developers may conceive of a voice assistant, such as Apple's Siri. Crucially, these examples are not simply AI in the sense of algorithmic intelligence and cognitive processes conducted by machines, i.e., machine learning [11]. They are also not merely artificial agents that can perceive an environment through sensors and respond accordingly with actuators [11]. These AI are embodied, embedded, and experienced within social contexts [6], during which interactions with people are a requirement.

Yet, we cannot say with certainty that human-AI interaction (HAI) contexts lead to the embodied AI being *socially* embodied, at least not all the time, in all contexts and cases, for all people. A human interface for AI on its own does not necessarily make the AI social or agentic [1,43], even when the AI may spark social behaviors in the people with whom it is interacting [19,21]. Work with children has shown that preexisting impressions and experiences with AI can heavily influence future experiences with other AI [42]. Explorations of co-creation—by definition, a social activity that makes use of forms of intelligence—with AI assistants or collaborators also reveal that sociality is elusive and con-

* Work completed while Postdoctoral Researcher at RIKEN AIP.

textual [23,29]. To illustrate, an industrial robot that normally carries out routine procedures on the assembly line can be programmed to give a high five to attendees at an office party. But not everyone who experiences the high five—bystanders, initiator, or even the programmer—will feel that the situation has crossed a social threshold. All of this points to a dynamic phenomenon that emerges from a certain combination of factors that we can recognize but as yet do not have a way of meaningfully describing or defining.

In this preliminary work, we argue that a new category[1] is needed to characterize such experiences. Specifically, we need a category that can represent this social phenomenon such that it can be applied to a variety of systems, contexts, use cases, and user experiences in a dynamic way. We propose to call this category "socially embodied AI." The rest of this paper serves to justify this positioning. To this end, we took a three-part approach. First, we developed a conceptual foundation by drawing on the relevant historical terms, models, and theories from technical and social fields in a literature review. This led us to a working definition for socially embodied AI. We then sought to test and validate this working definition through a card sort with experts in human-robot interaction and human-computer interaction who we did not make aware of our purpose. Our findings from this exercise validated the definition. We then used this definition, other findings from the card sort, and the foundational concepts from our literature review to construct a working ontology of socially embodied AI from human perception and social embodiment perspectives. In constructing this ontology, we were able to isolate a heuristic—what we propose to call the Tepper line—to demarcate the threshold between embodied AI and socially embodied AI. We then tested the ability of this ontology to faithfully describe the socially embodied AI phenomenon and illustrate the Tepper line heuristic through two case studies: a social robot and Siri. We end this paper with suggested next steps.

## 2 BACKGROUND: FROM THEORY TOWARDS "SOCIALLY EMBODIED AI"

We recognize that creating terms and ontologies are a "chicken or the egg" task: do we start with an idea and then justify it, or do we allow an idea to emerge naturally through the process of defining it? As Noy and McGuinness [28] suggest, we do both in an iterative fashion. We draw from the literature and then ground what we find in our own ideas, and later in the ideas of others. To start, we present our literature review.

Central to our concept is the notion of *embodiment*. Physical embodiment or form factor [40], which we will call the "body" for simplicity, may vary greatly. Some are standalone, like many social robots with their robotic bodies, and some require one or more principal technologies, such as chatbots, which are typically rendered in a browser accessed through a physical device, like a laptop computer. Importantly, none are programs, algorithms, processes, or abstract systems that have no direct contact with the world [33]. In contrast, the AI that power social media are not applicable, because they are not embodied and they are not social. Rather, they are hidden processes that facilitate social interactions among people and non-social interactions among non-people, like social bots [9].

Yet, interacting with people is not sufficient for *social* interaction. An example is the recommender system. Amazon's industry-standard product recommender [38], for instance, presents options to the end-user based on mined data of their previous purchases, page views, and so on. The user clicks or ignores these options, perhaps racking up page views for certain items. In this way, through a series of somewhat ambiguous interactions, the options "offered" by the AI to the user are refined over time. But few would describe this as a social activity.

Moreover, a *socially interactive* AI that is physically embodied may not constitute *social embodiment*, as originally conceived by social psychologists. According to Barsalou *et al.* [2], social embodiment is an explanation of how embodied states, such as gestures and facial expressions, are a *two-way street*. On the one side is social information processing (stimulus view), and on the other is social enactment (reaction view). This is a person-centered view of experience. As such, creating AI that provide certain social stimuli and reactions, mimicking human beings, may not be a solution. As Guzman [17] and Gratch [14] warn, we need to be careful when applying human models to interactions with machines, as they may not hold true. Further, we are still discovering what cues, properties, and situations trigger what social phenomena among human beings, let alone humans and non-human agents [22].

[1] We use "category" and "class" interchangeably. From an ontological perspective, "class" is more appropriate, but "category" may be clearer for the general reader.

Even so, theoretical and conceptual work suggest a path forward. Decades of work has shown that people tend to treat computers as people, unthinkingly and without realizing it [20]. The basic premise of the Computers Are Social Actors (CASA) paradigm is that people apply social roles, expectations, attitudes, and behaviors to computers that cue human embodiment. For instance, when gender markers were present, people rated their perceptions of a conversational AI's emotional intelligence in gender stereotyped ways [5]. By perceiving computers as people and reacting to these perceptions while interacting with computers, people are effectively denoting computers as social actors. This is, however, a one-way view that tells us more about the person than the AI or their relationship. Nevertheless, it seems that people are primed to interact with embodied AI socially.

Similar notions have been proposed in social robotics. Dautenhahn, Ogden, and Quick [8] linked embodiment to social embeddedness, or the robot being situated within and able to (re)act to a social environment that includes humans, They took a robot-centered perspective, focusing on the robot's ability to perceive and interact with a human. Their framing did not consider the human's perceptions of the robot, social or otherwise. Dautenhahn went on to propose four key characteristics of social robots [7]: they must be socially evocative (anthropomorphizable), socially situated, sociable or driven to be social, and socially intelligent in a human-like way. Miller and Feil-Seifer [24] later proposed three properties for social robots: morphology (the robot's physical form, effectively an anthropomorphic measure similar to Dautenhahn's "socially evocative" criteria), situatedness (in the same sense that Dautenhahn meant by "socially situated"), and embodiment (able to interact using its physical form in the environment in which it is situated, particularly in this case socially with people). Fong *et al.* [10] proposed that robots designed primarily for a social purpose should be distinguished as *socially interactive robots*.

Much can be harnessed from the social robotics space, but we are aiming for brevity and technology neutrality. Robots, chatbots, voice assistants, recommender systems, chatbots, virtual humans, and other human-interfacing technologies that could be socially embodied are *agents*. Put simply, an agent is an entity that can act in some world, physical or virtual [12]. Agents may follow their own will or agenda and thus be autonomous, or they may follow preprogrammed actions or scripts. But agents do not necessarily interact with others (humans or otherwise), even when they appear humanlike or social. Guzman and Lewis [17] pinpointed the common thread between human-facing agents that have or use human or social cues: the AI. Grounded in Guzman's [16] preliminary Human-Machine Communication (HMC) framework, they argued that computers are no longer simply mediators of interaction, but participants. They suggested three perspectives: functional (what the AI does and how people feel about it), relational (a longer-term, contextual view to interactions with such AI, in line with social embodiment as defined by Barsalou *et al.*), and metaphysical (the ethics and future of communication that is not merely the domain of humans, or a way to define "human"). These perspectives are drawn from current research trajectories that call into question the previously held notion of communication as a *human* process. Practically, the use of AI-based agents in communication technologies is leading to a reality where machines are becoming communicators, not just facilitators of communication between people. Any technology that provides us with the sense that we are interacting with another individual will encourage us to assign social dimensions to that exchange.

What is generally meant by AI is non-human, yet human-engineered intelligence. Still, a standard definition has long eluded not only the CHI and HAI communities [43] but also those in the field of AI [27]. Nilsson, one of the field's co-founders, suggested that people must *perceive the existence of intelligence*. He defines intelligence as "a quality that enables an entity [agent] to function appropriately and with foresight in its environment" [27:13]. This is a general view of intelligence. Social intelligence relates to skill in social interactions [41]. It too is difficult to define. One perspective that matches Barsalou *et al.* [2] is by Goleman [13]: social awareness (e.g., empathy, social cognition) as *input* and social facility (e.g., concern, self-presentation) as *output*. As with Nilsson and the CASA model, it is less about the AI actually doing the work than it is about the AI being *perceived by people* as doing the work. In line with theory of mind, we assume a "mind" in others that is comprised of what they know, expect, and experience [34]. In effect, we are biased to see social intelligence even when there is none. For AI, social intelligence is constructed through HAI experiences rather than being a product of engineering social intelligence in the AI.

Taken together, we propose the term "socially embodied AI" and define it as follows:

> The dynamic state that an embodied AI attains when it is perceived by a person as a social agent during interaction. It can apply regardless of form factor, morphology or "body," actual intelligence, intentions of the creators, and perceptions of others. As a state, it may be lost or regained depending on the stability of the factors leading the person to perceive social embodiment. Hence, it is a circumstantial category.

Next, we put it to the test through a card sort workshop with experts.

## 3 CARD SORT: ENHANCING WITH EXPERT INSIGHTS

Our review of the literature provided 71 descriptors and concepts for defining AI in various ways. To validate the term and concept externally, we conducted an open card sort with experts: knowledgeable others naïve to our idea.

### 3.1 Methods

A card sort is a categorization process where some kind of information is placed on cards that are then grouped by people in some way [39]. The outcome is a form of information architecture: a representation of the mental models that people have about that information. In this way, card sorting makes tacit information explicit. It has traditionally been used in end-user system design and user testing (e.g., [30,31]), but can also be used as a knowledge elicitation technique for ontological and conceptual modelling [18,35]. We provided an initial set of 71 cards, each featuring one basic descriptor. These were extracted by the first author from the definitions, features, and types of technologies described in the papers from our literature review. We also provided three category cards on the major application areas in AI: industrial, medical, and social. We conducted an open card sort, encouraging participants to add to and disagree with our choices. Our invitation was framed as "defining and categorizing robots," but in the workshop we asked participants to consider other technologies, morphologies, and types of AI, and how they fit together. We did this to avoid leading participants directly to our own conclusions while providing the same foundations and framing. This balancing act was meant to limit confirmation bias in our results [26].

### 3.2 Participants

We recruited six experts (all male, aged 22-45) who were external to our institution. All had an undergraduate degree in a relevant field, e.g., computer science, and were working with AI in some capacity. They were recruited through our local network of peers and collaborators. They were not aware of our socially embodied AI concept or aspirations. They were not compensated for participation except for snacks and drinks during the workshop.

### 3.3 Procedure

Participants were introduced to card sorting in a short presentation. The cards were then laid out randomly on the table. Participants were asked to add to, modify, and categorize the cards in two rounds: first, to create a general definition of "robot," and then with respect to the three major categories (industrial, medical, social). They were encouraged to sort the cards as made sense to them and discuss and negotiate with each other when disagreements occurred. They were allowed to re/label until they were satisfied. They also created individual definitions before and after the workshop. We used these to account for group influence and to validate our conceptual framework.

### 3.4 Results and Discussion

Supplementary Materials Figure 1 illustrates the card sort results. Participants generated three descriptor cards, three category cards, and a cardinal center. In the first round, a set of "basic functions" was created based on a historical perspective of robot applications: the desire to industrialize, routinize, and standardize the production of goods and services. Robots were seen as an extension to the factory line system, a means to offload menial labour from human workers. This view has infiltrated other domains, particularly medical. But while such domains may involve physical and repetitive human labour or a standard of care, they are also firmly entrenched in social systems, roles, and relationships. This is reflected in the second round, where participants created a cardinal center with *factory–home* and *social–non-social* axes. This reflects the historical foundations of institutional (e.g., factory, hospital, school) and personal (e.g., home, community, culture) spaces against the social, human-derived features of robots. Participants treated non-humanoid robot descriptors as basic, traditional, and non-social, while humanoid ones were treated as social. This distinction reveals an unwritten meaning: *what is human is social.*

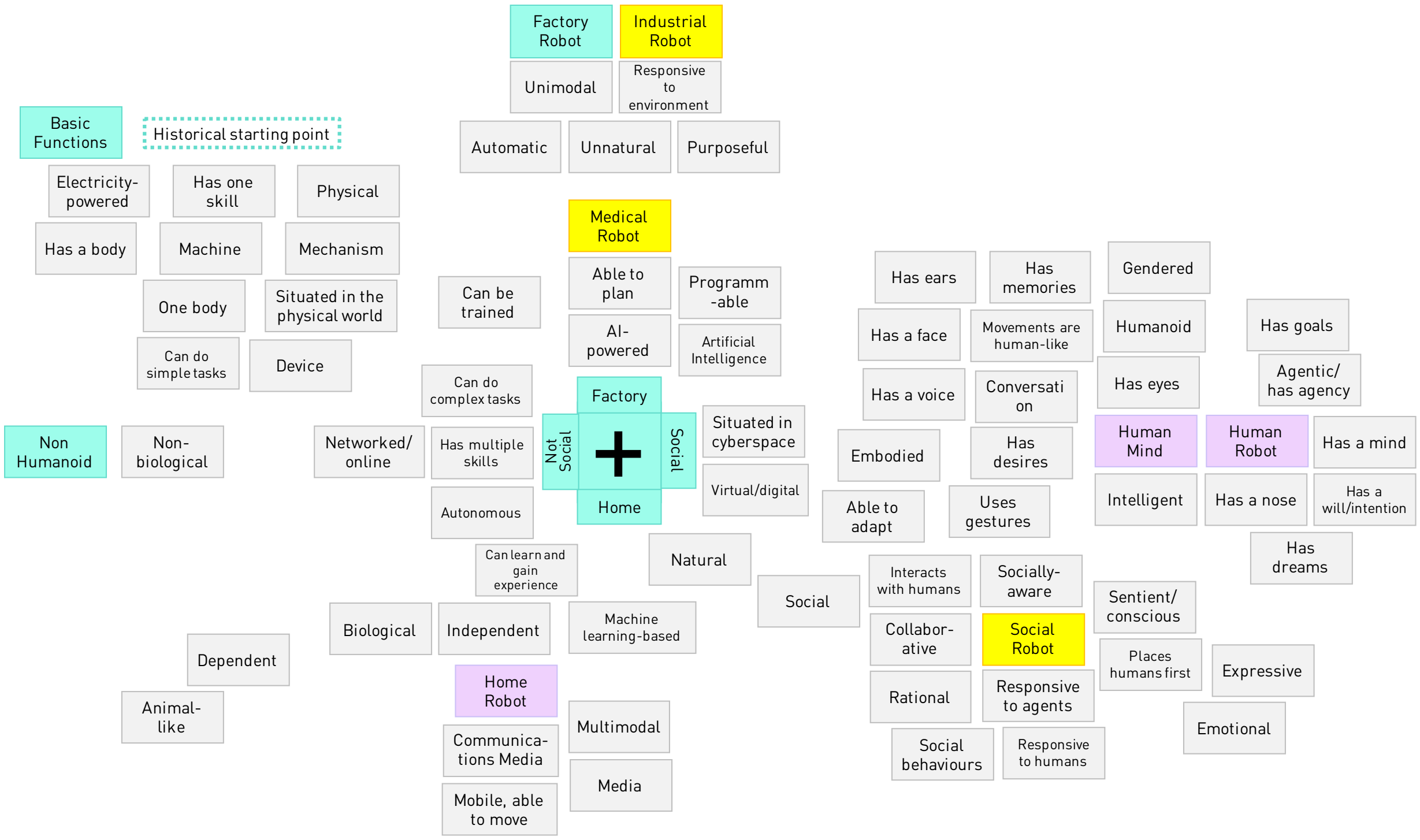

**Supplementary Materials Figure 1: The card sort results. Predefined cards included grey for the descriptor cards and yellow for categories. Participant-created cards included purple for descriptor cards and cyan for categories.**

Still, the creation of the "human mind" and "human robot" categories, and the placement of such features as "has eyes," "has goals" and "gendered" within it, separate from the "social robot" category, suggests that *human-like* is not necessarily *social*, i.e., inter/actions are not necessarily social. Similarly, "AI" and related terms are placed around the center, suggesting that AI is not necessarily social, or human. Perception of sociality and the nature of the situation may be key. The gaps in the north-east and south-west areas—*social-industrial* and *non-humanoid-home*—establish a link between domain and sociality. Agents in the non-social factory and industrial domains are not expected to have social features. This reinforces the idea that the human and the social are connected. We next used these finer grained details to refine the concept of socially embodied AI into an initial ontology.

## 4 TOWARDS AN ONTOLOGY OF SOCIALLY EMBODIED AI

In taking on the task of proposing an ontology, we recognize the importance of being comprehensive. By this we mean basing the concept and term on consensus, wherever possible, and being self-critical and iterative [28] to achieve rigor. Here, we lay a groundwork that unites existing perspectives in a meaningful way.

The working ontology is presented in Figure 1a. For comparison, an embodied AI version without the Social class is presented in Figure 1b. Person and Embodied AI are subclasses of the Participant class, in line with view that machines must graduate from mediator to participant [17]. Both participate in an Interaction. The Person has a Perception of the Interaction that is Social. The Social Perception is encouraged and enforced by the Re/action, Morphology, and Intelligence properties of the Embodied AI [2,24]. The Interaction is part of a Situation [7,10,24], which has, at minimum, a Context and a Purpose. These may not be explicitly designed to be Social but must be perceived by the Person as Social [2,13,27]. The Social Perception of the Interaction drives these perceptions. The resulting socially perceived Situation then begets the Social Embodiment property of the Embodied AI. Thus, the Social class cascades through the Perception, Interaction, and Situation classes to result in Social Embodiment.

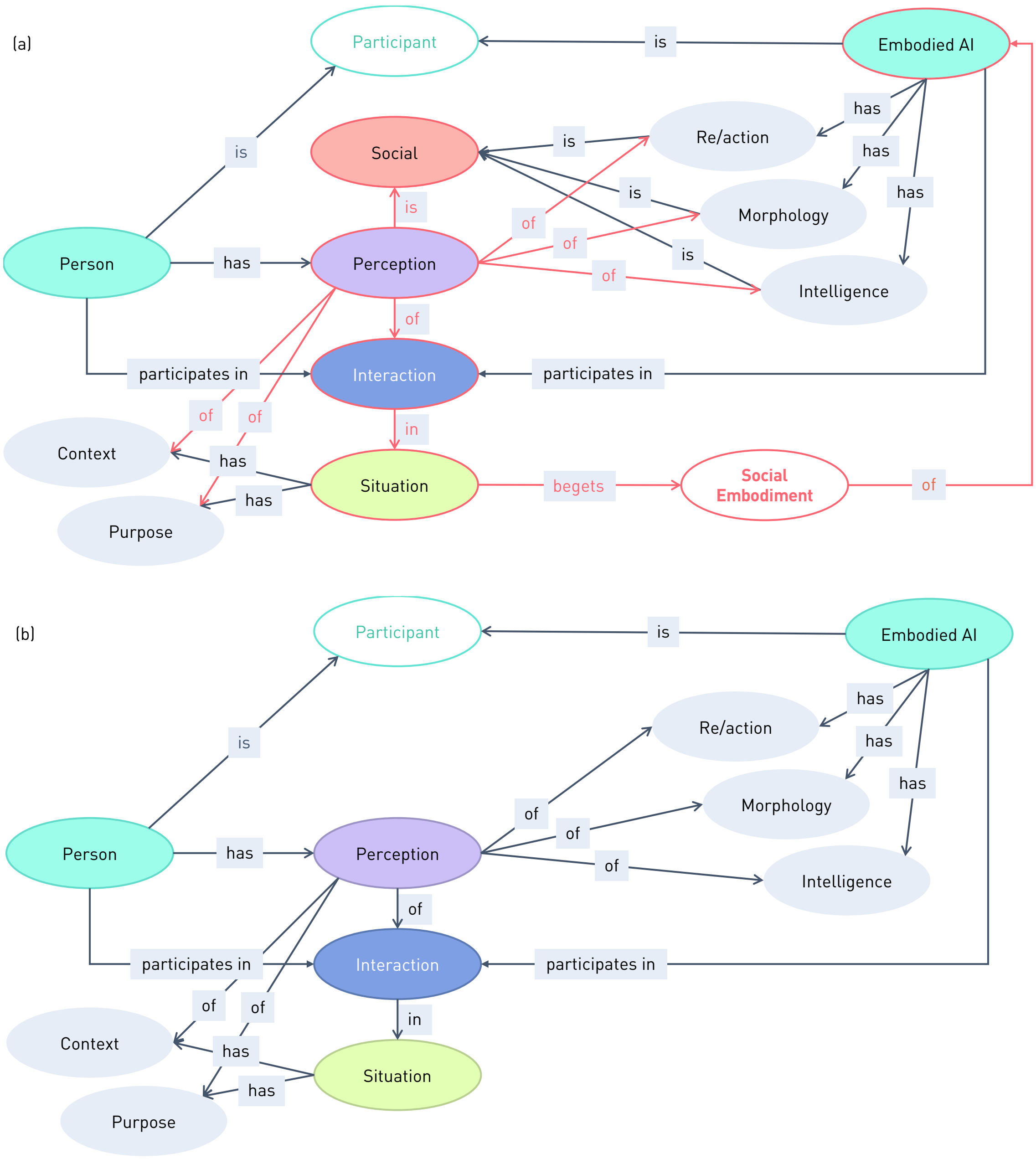

**Figure 1: (a) The working socially embodied AI ontology (top). (b) The general embodied AI version (bottom)**

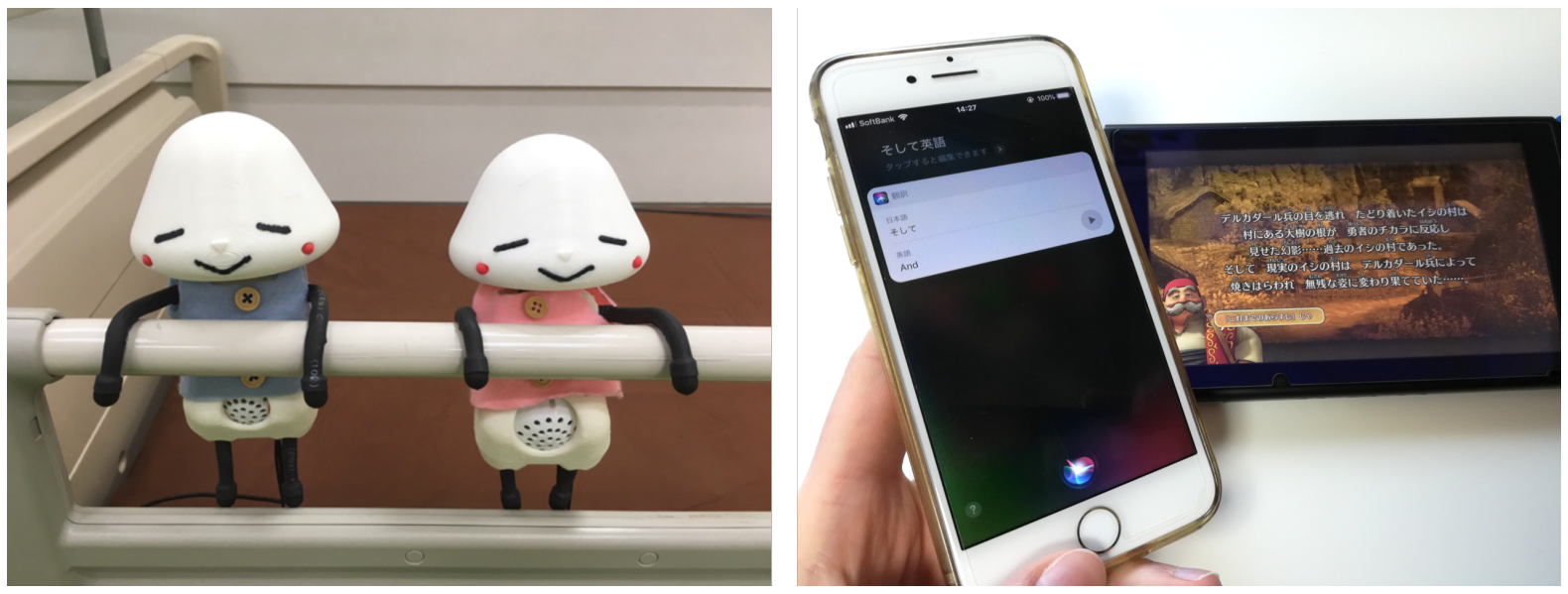

**Figure 2: (a) Two SideBot robots (left). Apple Siri and the Nintendo Switch (right).**

In summary, the Embodied AI becomes a Socially Embodied AI when these criteria are met:

- the Embodied AI has characteristics that can be perceived by a Person as Social markers [20,24];
- the Interaction occurs in a Situation where the Context and Purpose are perceived as Social [2];
- the Person perceives the Embodied AI, the Interaction, and the Situation as Social [17,20].

Figure 1b depicts the case when the Social class is absent. Importantly, a Person can still Interact with an Embodied AI that is Intelligent and so on, but it is not Socially Embodied. Additionally, the ontology takes no stance on the type of Embodied AI. Social robots, voice assistants, conversational agents, and more can qualify as Socially Embodied AI if these conditions are met. Thus, the ontology is flexible, representative, and comprehensive. It allows us to classify a specific phenomenon without binding us to a certain category, field of study, or disciplinary tradition.

## 5 DELIMITING THE THRESHOLD: THE TEPPER LINE

"Socially Embodied AI" is a circumstantial category that is broadly applied in a dynamic way. The classes and their properties are high-level, allowing for flexibility, but also imprecision. A question then emerges: When or under what circumstances does an embodied AI cross the social line? At a low level, this will differ person-by-person and context-by-context, as well as with the strength of the embodied AI in terms of its design, technical robustness, and so on. At a high level, we will need to apply the ontology to existing technologies as well as run new studies that test the bounds of the line between mere embodied AI and socially embodied AI. We call this line the "Tepper" line after Sheri S. Tepper, who was a science fiction author known for criticizing, playing with, and upending social expectations and assumptions about agency and communication through her work [3]. She explored how social perceptions modulated her characters'—and the reader's—ability to distinguish *object* from *being*, from "lower" lifeform to the anthropomorphic standard. Unlike other thinkers within the imagined futures space, Tepper was *explicitly social* in her view of intelligence as a gateway for humans to recognize non-human beings as equals. As such, we named the threshold across which an AI-based agent becomes socially embodied after her. We next put our working ontology to the test by considering its ecological validity [4] using two example cases.

### 5.1 First Case: SideBot

SideBot (Figure 2, left) is a long-term use personal robot designed to prevent falls and provide companionship to bedridden patients at care facilities and hospitals [25]. Styled after Japanese Kokeshi dolls, it uses natural language to communicate and is connected to a smart mattress that can detect posture and movements. In emergency situations, it can use conversation to discourage the patient from getting out of bed until a care worker arrives. It can also engage the patient in trivia games and play songs. As a long-term companion, SideBot has the potential to become a partner for the patient rather than merely a telehealth tool for care workers. In a pilot study [25], SideBot was placed in older adults' homes for one week. Here, we consider the case of one participant, who lived alone but was active socially. At first, while she liked the shape of the face and its expression (*Morphology*), she felt its presence in her home was awkward. Yet, after a week, she expressed her attachment to the robot, saying that she would "feel lonely" after it was removed (*Purpose*). She explained that SideBot talks "to" her and plays songs "for" her (*Intelligence* and *Re/actions*), and is "something of a true companion," despite its limitations. After SideBot was removed, she wanted it to "come home again" (*Context*). In this way, SideBot shows where the Social Perception of the Embodied AI properties mediated a change in the Social Perception of the Situation that then led to it being perceived as socially embodied. Even so, SideBot is not yet sustainable as a companion. Its communication mechanisms are "hard-wired" and sequentially, if autonomously, deployed. This means that the robot often goes through the same repertoire of responses in the same sequence—repetitive and unrealistic. Still, its embodiment makes it possible to treat it as a companion despite its imperfect capabilities. Indeed, SideBot illustrates how an embodied AI's status as socially embodied is a two-way interaction between human and the agent.

### 5.2 Second Case: Siri

Siri is a voice-based virtual personal assistant designed by Apple and available through Apple OS. Siri has a natural language user interface and primarily acts as a voice-based search tool. Yet, as some have started to ask [15], Siri has the potential to become more than a tool, even a partner. Siri represents a case where the Situation plays a key role in its transition to a socially embodied AI. This example focuses on the situation of the first author residing in a country where the official language is not her native language. She is using several methods to gain proficiency. She is also a gamer. Knowing that video games can provide an immersive, motivational, and challenging space to learn a language (see e.g., [32]), she decided to purchase a Nintendo Switch with the game Dragon Quest XI set to the country's language. Since the game uses a combination of text and voice acting with subtitles, it is possible to hear and "read" the words used without knowing the meaning. Yet understanding the meaning of the words is essential for making progress through the game, or even to use the interface. Enter Siri (Figure 2, right). The first author started to use Siri in an "as-needed" fashion as a dictionary, looking up words only when she did not know the meaning. Siri thus satisfied the general HAI experience represented in Figure 2b, without the Social property. But as time went on, she found herself oscillating in her behavior and intentions towards Siri. At times, she found herself treating Siri as a partner: a source of help outside of the game (*Purpose*), perhaps more knowledgeable, like a friend who had played the game before and just wanted to watch this time. Over time, Siri's typical clipped responses (*Morphology*) and occasional misfires (*Intelligence*) began to feel more like an older sibling who was not always paying attention (*Context*), curtly providing just the essentials for her to work it out on her own (*Purpose*, *Re/action*). Still, like SideBot, Siri is too limited to maintain a suspension of disbelief. For instance, it is not possible to ask Siri to retrieve another definition for a term with the same reading: it will repeat the "optimal" definition, disrupting the Social Perception of its Re/action property. Yet, the first author could move on and enter a new Situation in which Siri responded in a meaningful way, re-crafting the magic circle [37] despite any earlier hiccups. In this way, Siri shows how an AI's status as socially embodied is circumstantial, flawed, and yet recoverable.

## 6 CONCLUSIONS

We have captured the dynamic sociality of embodied AI through a literature-informed ontology that was validated in a card sort exercise with experts. In two case studies, we have highlighted a threshold—the Tepper Line—to demarcate embodied and socially embodied AI. Future work will need to put the ontology to the test, applying it to other embodiments, searching out outliers and extreme cases, and testing its bounds. We also only involved experts in robotics and AI for our card sort. Future work can further validate and test the socially embodied AI concept with non-experts or a mix of lay people and experts. We hope to show how such an ontology can bring together diverse AI-based agents in a way that is dynamic, situated, and grounded in the social perceptions of the human user.

## ACKNOWLEDGMENTS

We sincerely thank our reviewers. Funded in part by the Japanese Society for the Promotion of Science (JSPS) Grants-in-Aid for Scientific Research (KAKENHI #20H05022 and #20H05574) and AMED grant # JP19he2002014.